# Thermoelectric Properties of Scaled Silicon Nanowires Using the sp$^3$d$^5$s*-SO Atomistic Tight-Binding Model and Boltzmann Transport

Neophytos Neophytou and Hans Kosina

Institute for Microelectronics, TU Wien, Gußhausstraße 27-29/E360, A-1040 Wien, Austria

e-mail: {neophytou|kosina}@iue.tuwien.ac.at

## Abstract

As a result of suppressed phonon conduction, large improvements of the thermoelectric figure of merit, *ZT*, have been recently reported for nanostructures compared to the raw materials' *ZT* values. It has also been suggested that low dimensionality can improve a device's power factor as well, offering a further enhancement. In this work the atomistic sp$^3$d$^5$s*-spin-orbit-coupled tight-binding model is used to calculate the electronic structure of silicon nanowires (NWs). The linearized Boltzmann transport theory is applied, including all relevant scattering mechanisms, to calculate the electrical conductivity, the Seebeck coefficient, and the thermoelectric power factor. We examine n-type nanowires of diameters of 3nm and 12nm, in [100], [110], and [111] transport orientations at different carrier concentrations. Using experimental values for the lattice thermal conductivity in nanowires, the expected *ZT* value is computed. We find that at room temperature, although scaling the diameter below 7nm can be beneficial to the power factor due to banstructure changes alone, at those dimensions enhanced phonon and surface roughness scattering degrades the conductivity and reduces the power factor.

**Index terms:** thermoelectric, conductivity, tight-binding, atomistic, sp$^3$d$^5$s*, Boltzmann transport, Landauer, Seebeck coefficient, silicon, nanowire, *ZT*.

# I. Introduction

The ability of a material to convert heat into electricity is measured by the dimensionless figure of merit $ZT=\sigma S^2 T/(k_e+k_l)$, where $\sigma$ is the electrical conductivity, $S$ is the Seebeck coefficient, and $k_e$ and $k_l$ are the electronic and lattice part of the thermal conductivity, respectively. The recent progress in the synthesis of nanomaterials allows the realization of low-dimensional thermoelectric devices based on one-dimensional (1D) nanowires (NWs), and two-dimensional (2D) thin films and superlattices [1, 2, 3, 4]. Low-dimensional materials offer the possibility of improved thermoelectric performance due to the additional length scale degree of freedom in engineering $S$, $\sigma$, and $k_l$ through partial control over the dispersions and scattering mechanisms of both electrons and phonons. Higher $ZT$ values in nanostructures compared to their bulk material's values can therefore be achieved [1, 2]. Enhanced performance was recently demonstrated for silicon NWs [1, 2]. Although bulk silicon has a $ZT_{bulk} \sim 0.01$, the $ZT$ of silicon NWs was experimentally demonstrated to be $ZT\sim0.5$. Most of this improvement has been a result of suppressed phonon conduction ($k_l$), but low dimensionality can be beneficial for increasing the power factor ($\sigma S^2$) of the device as well [5, 6, 7]. The sharp features in the low-dimensional density of states ($g_{1D}(E)$) as a function of energy can improve $S$ [5, 6, 7], as this quantity is proportional to the energy derivative of $g_{1D}(E)$. Carrier confinement through structural scaling in certain directions can also potentially improve $\sigma$ through reduction of the material's effective mass [8, 9]. Of course an improvement in $\sigma$ can degrade $S$, since these two quantities are inversely related. At the nanoscale, however, subband engineering techniques can be used to optimize this interrelation and maximize the power factor and further improve $ZT$.

In this work, we calculate the electrical conductivity, the Seebeck coefficient, and the electronic part of thermal conductivity of thin silicon NWs using the sp$^3$d$^5$s*-spin-orbit-coupled atomistic tight-binding model. The model provides an accurate estimate of the electronic structure, while being computationally affordable. We examine cylindrical n-type NWs, i) of diameters $D$=3nm (ultra scaled) and $D$=12nm (electrically approaching bulk), ii) in [100], [110] and [111] transport orientations, and iii) different carrier concentrations. The thermoelectric coefficients of interest ($\sigma$, $S$ and $k_e$), are calculated within the linear transport Boltzmann formalism [10, 11] including all relevant scattering mechanisms. Using experimentally measured lattice thermal conductivity values, the $ZT$ values of the NWs are estimated.

We find that at room temperature, although from bandstructure considerations alone the power factor ($\sigma S^2$) and $ZT$ can be ideally improved for NW diameters smaller than ~7nm [12], at these diameter scales phonon scattering, and especially surface roughness scattering (SRS), become increasingly important and significantly reduce the power factor. Improved power factors in n-type Si NWs are, therefore, not obtained with diameter reduction, and are rather dependent on the value of the conductivity, contrary to calculations performed using ballistic transport, which pronounce the improvement of the Seebeck coefficient and predict power factor improvements. For power factor optimization, we show that as the diameter scales down to $D$=3nm, [110] NWs seems to be the optimal choice (although the orientation dependence is rather weak). The peak of the power factor appears at carrier concentration levels of ~$10^{19}$/cm$^3$. When this is achieved by doping, at such high concentrations, ionized dopants dominate the scattering processes and strongly affect the power factor.

## II. Approach

The NWs' bandstructure is calculated using the 20 orbital atomistic tight-binding spin-orbit-coupled model (sp$^3$d$^5$s*-SO) [13]. Each atom in the NW is described by 20 orbitals (including spin-orbit-coupling). The NW description is built on the actual diamond lattice and each atom is properly accounted in the calculation. It accurately captures the electronic structure and the respective carrier velocities, and inherently includes the effects of quantization and different orientations. The sp$^3$d$^5$s*-SO model was extensively used in the calculation of the electronic properties of nanostructures with excellent agreement to experimental observations on various occasions [14, 15]. Details of the model are provided in [8, 13, 14, 15]. We consider here infinitely long, cylindrical silicon NWs. The electronic structure of ultra scaled devices is sensitive to diameter and orientation [8, 9, 16]. Differences in the shapes of the dispersions between wires of different diameters and orientations, in the number of subbands, as well as the relative differences in their placement in energy, can result is different electronic properties. We, therefore, consider three different transport orientations [100], [110], and [111] and two different diameters $D$=3nm and $D$=12nm.

The linearized Boltzmann formalism is used to extract the thermoelectric coefficients for each wire using its dispersion relation and all relevant scattering mechanisms. In this method the intermediate transport distribution function is defined as [10, 11]:

$$\Xi(E) = \sum_{k_x,n} v_n^2(k_x) \tau_n(k_x) \delta(E - E_n(k_x))$$
$$= \sum_n v_n^2(E) \tau_n(E) \frac{g_{1D}^n(E)}{A}. \quad (1)$$

where $v_n(E) = \frac{1}{\hbar} \frac{\partial E_n}{\partial k_x}$ is the bandstructure velocity, $\tau_n(k_x)$ is the *velocity* relaxation time for a state in a $k_x$-point and subband $n$, $A$ is the NW's cross sectional area and

$$g_{1D}^n(E_n) = \frac{1}{2\pi\hbar} \frac{1}{v_n(E)} \quad (2)$$

is the density of states for 1D subbands (per spin). From this, the electrical conductivity ($\sigma$), the Seebeck coefficient ($S$), and the electronic part of the thermal conductivity ($k_e$) can be defined as:

$$\sigma = q_0^2 \int_{E_C}^{\infty} dE \left(-\frac{\partial f_0}{\partial E}\right) \Xi(E), \quad (3a)$$

$$S = \frac{q_0 k_B}{\sigma} \int_{E_C}^{\infty} dE \left(-\frac{\partial f_0}{\partial E}\right) \Xi(\varepsilon) \left(\frac{E-\mu}{k_B T}\right), \quad (3b)$$

$$\kappa_0 = k_B^2 T \int_{E_C}^{\infty} dE \left(-\frac{\partial f_0}{\partial E}\right) \Xi(E) \left(\frac{E-\mu}{k_B T}\right)^2, \quad (3c)$$

$$\kappa_e = \kappa_0 - T\sigma S^2. \quad (3d)$$

Figure 1 shows the electronic structure of a cylindrical NW in the [110] transport orientation. There are three two-fold degenerate valleys in the dispersion relation, one placed at the Γ point, and two placed off-Γ. The dispersion relations are obtained by diagonalizing the atomistic Hamiltonian. For every $k_x$-state at a subband $n$, the carrier velocity and the density of states are extracted. The scattering rates for every state are also extracted directly from the dispersions using Fermi's Golden rule [17, 18]. Elastic and inelastic scattering processes are included as indicated in Fig. 1. Acoustic and optical phonons (with parameters from Ref. [17] with the exception of $D_{ADP}^{electrons} = 9.5$ eV [19]), surface roughness scattering, and impurity scattering are included. The matrix element for surface roughness scattering is extracted using an $\Delta_{rms}$ value and an autocorrelation length ($L_C$) for the roughness taken from [18]. The scattering strength is derived from the shift in the band edges with quantization (for each valley separately). This is a simplified way of treatment of SRS, but it is a valid approximation for ultra scaled channels, where the SRS limited low-field mobility follows a $D^6$ behavior, where $D$ is the diameter of the channel, originating from quantization [18, 20, 21]. The elastic and inelastic scattering processes considered (including both *f*- and *g*-processes for all six relevant phonon modes in Si), are treated using the *bulk* Si selection rules. For example, each valley in Fig. 2 is two-fold degenerate, but only intravalley scattering is allowed, i.e. each valley scatters only within itself. That is why inelastic processes are allowed not only between, but also within the Γ and off-Γ valleys as shown in Fig. 1. Finally, the atomistically extracted wavefunctions are used to calculate the overlap integrals (form factors) between the initial and final states in the calculation of the scattering matrix element. A detailed description of the scattering treatment will be presented elsewhere.

Our simulations include up to ~5500 atoms (for the $D$=12nm NWs). To be able to extend the computational domain to such scale, certain approximations are employed. The most important ones are: i) We use bulk instead of confined phonons. As discussed in Ref. [22], the effect of confined phonons on the conductivity is not large, and does not affect the basic trends of the results. ii) We use bulk deformation potential scattering parameters. The values for the deformation potentials for nanostructures somewhat vary in the literature [18, 19] and different values can indeed have a strong influence on the conductivity [22]. iii) Structure/surface relaxation is not considered in this study. This is subject to the surface passivation type. It might be important for the absolute performance numbers in the smaller $D$=3nm NWs. Admittedly, these assumptions and especially assumption (ii) can have an influence on the magnitude of our results and this needs to be investigated in coordination with experimental data, which at this point are sparse. We believe, however, that the dimensionality trends we describe are independent of these assumptions. The basic trends and findings with respect to the importance of the conductivity versus the Seebeck coefficient in the power factor, or the different insight gained from ballistic versus scattering based transport will not be altered.

## III. Results and Discussion

Figure 2a shows the Seebeck coefficient of NWs in the [100] (blue), [110] (red) and [111] (green) transport orientations, for $D$=3nm (solid) and $D$=12nm (dashed) as a function of the carrier concentration. To investigate the thermoelectric properties of perfect NWs, in this case we only consider the effect of phonon scattering. SRS and impurity scattering will be considered further on. At $D$=3nm some orientation dependence is observed. The [111] NW with six-fold degenerate valleys has an advantage over the other NWs, whereas the [110] NW with a two-fold degenerate lowest valley has the lowest Seebeck coefficient. The $D$=12nm NWs have only minor orientation dependence. They all have lower Seebeck coefficients than the $D$=3nm NWs, which points to the initial driving reason for investigating the thermoelectric performance of pure 1D nanostructures [5, 6]. Although the differences between the different diameters and orientations are small, some optimization directions can be identified. The electrical conductivity in Fig. 2b, on the other hand, follows the inverse trend. The conductivity of the $D$=3nm NWs is degraded compared to the conductivity of the $D$ = 12 nm NWs. The reasons for this are: i) the stronger phonon scattering as the diameter is reduced (originating from larger wavefunction overlaps between the states), and ii) the fact that the number of subbands that participate in transport scales less than linearly with diameter scaling. At the same carrier concentration, the latter pushes the band edges away from the Fermi level, shifting the conductivity and Seebeck coefficient to the right in Fig. 2. It is interesting to notice the order of the conductivity in Fig. 2b. For the larger diameter NWs, the [100] NWs are the best, followed by the [111] and afterwards by the [110] NWs. For the smaller diameter NWs, the [110] NW is the one with the highest conductivity, followed by the [100] and finally by the [111] NW. The reason that the [110] NW overpasses the other two in the $D$=3nm case is that for extremely scaled NWs the effective mass of the [110] NW is reduced, whereas the masses of the other two NWs are increased [8].

As the NW diameter reduces from $D$=12nm down to $D$=3nm, additionally to phonon scattering, SRS also increases. These two scattering mechanisms greatly reduce the conductivity of the smaller diameter NWs. To illustrate the effect of stronger phonon and SRS mechanisms for the smaller NWs, in Fig. 3 we plot the power factor for the [111] NW for $D$=12nm (dashed) and $D$=3nm (solid). The arrow shows the direction of diameter decrease. Figure 3a shows the devices' power factor per unit area under *ballistic* transport conditions, extracted using the Landauer formalism [12, 23, 24]. Clearly, as the diameter reduces, the power factor increases. Our previous work, using only ballistic transport considerations [12], showed that this increase as well as NW orientation performance differences are linked to improvements in the Seebeck coefficient. This originates from the different subband degeneracies of the various electronic structures that determine the position of the Fermi level at a given carrier concentration. Figure 3b shows the same result for simulations in which only phonon scattering is considered. In this case the performance of the two NWs is somewhat more similar, which means that the $D$=3nm NW is affected more by scattering. (Note that the units of the power factor are different in Fig. 3a and Fig. 3b since in the case of ballistic transport we compute the conductance-$G$ instead of conductivity-$\sigma$). Finally, Fig. 3c shows results for which phonons and SRS are considered. SRS has a strong negative effect on the conductivity of the $D$=3nm NW, thus significantly reducing its power factor, whereas it does not affect as significantly that of the $D$=12nm NW. We note here that the Seebeck coefficient is not affected much from case to case since it is independent of scattering at first order [24].

The variation in performance between Fig. 3a, 3b and 3c, therefore, originates from reduction in the electrical conductivity.

Figure 4 shows the power factor for the three NW orientations for $D$=12nm and $D$=3nm versus the carrier concentration (the same labeling as in Fig. 2 is used). In Fig. 4a phonon and SRS are considered. The power factors of the $D$=3nm NWs are reduced compared to the power factors of the $D$=12nm NWs regardless of orientation. It is evident from this that the effect of the Seebeck coefficient in influencing the power factor is minimal, whereas that of electrical conductivity is stronger. For the $D$=3nm NWs the [110] orientation has the best performance followed by the [100] and [111] orientations. The power factor peak of the [111] NW appears at much higher carrier concentration levels, which might not even be technologically feasible. For the larger diameters the [100] followed by the [111] NW exhibit the best performance, whereas the [110] one lacks behind. The orientation dependence is, however, small; only ~20% at most.

In Fig. 4a the peak of the power factor appears at carrier concentrations as large as $n_0$=10$^{19}$/cm$^3$. If such levels are achieved by impurity dopants in the channel, then as we know from bulk Si channels, impurity scattering will strongly degrade the conductivity. In Fig. 4b we plot the power factor versus carrier concentration, but in addition to phonons and SRS we also consider impurity scattering. We consider the number of impurities to be equal to the carrier concentration (charge neutrality in the channel). Indeed this scattering mechanism is dominant in Si NWs too, and at ~10$^{19}$/cm$^3$ it further degrades the power factor as shown in Fig. 4b (compared to Fig. 4a). The magnitude of the reduction can quite vary for different NW cases.

Recent works have reported that the thermal conductivity of Si NWs scaled down to 15nm or 20nm in diameter can be reduced by two orders of magnitude from its bulk material value, and can be as low as $k$=1-2W/mK [1, 25, 26]. This reduction is responsible for the enhancement in the thermoelectric figure of merit (*ZT*) of silicon NWs, which was measured to be close to unity. Using the measured value $k_l$=2W/mK for the $D$=15nm NW in Ref. [26], we estimate the expected *ZT* using the calculated power factors and $k_e$ for the NWs considered in this study. The *ZT* for the *D*=3nm and *D*=12nm in the [100], [110] and [111] NW orientations versus the carrier concentration are shown in Fig. 5 (with the same labeling as in Fig. 2 and Fig. 4). In Fig. 5a we only consider phonon and SRS. For the large diameter cases, *ZT* values close to or larger than unity can ideally be achieved. For the smaller diameters, the *ZT* is reduced following the power factor behavior. Once impurity scattering is also considered in Fig. 5b, the *ZT* drops compared to the phonon and SRS results of Fig. 5a following the trend of the power factors in Fig. 4. We would like to mention here that the value $k_l$=2W/mK used is measured for *D*=15nm Si NWs and might be smaller for smaller NW diameters, or even differ between orientations [27]. In such cases the *ZT* can even be higher. The *ZT* values provided in Fig. 5 are just estimates, but are in relative agreement to other reports in the literature, both theoretical [28] and experimental [1, 2].

## IV. Design Considerations

Comparing the magnitude of the power factor and *ZT* for the n-type *D*=3nm NW cases (Fig. 5), for carrier concentrations up to $n_0$=10$^{19}$/cm$^3$ the [110] NW with a two-fold degenerate Γ-valley and light effective mass (*m**=0.16m$_0$) (Fig. 1), has a larger *ZT* than the [100] NW with a four-fold degenerate Γ-valley but heavier effective mass (*m**=0.27m$_0$). The *ZT* of the [100] and [111] NWs overpass the *ZT* of the [110] NW only at higher doping concentrations. At larger diameters, when only phonons and SRS are considered, the [100] NW has the highest conductivity (four-fold degenerate light mass valleys of *m**=0.19m$_0$ and two-fold valleys of *m**=0.89m$_0$) and exhibits the highest *ZT*. The [110] NW has a two-fold degenerate valley of *m**=0.19m$_0$, but also a four-fold degenerate valley of *m**=0.55m$_0$, whereas the [111] NW has six-fold degenerate valleys of *m**=0.43m$_0$. Therefore, in either case, the *ZT* is higher in NWs with the largest conductivity instead of the largest Seebeck coefficient. We note that the purely ballistic transport results [12] indicate the reverse, namely that the NWs with the largest Seebeck coefficient exhibit the largest power factors and *ZT*. When impurity scattering is also considered, the conductivity of the [110] is less degraded than that of the other two orientations, and the power factor and ZT for this NW orientation is the highest.

For optimal results the electronic structure can be engineered using proper quantization and even strain. One needs to target: i) Increase $S$ by allowing more valleys nearby in energy, or using transport orientations with higher degenerate subbands as in the case of n-type [111] NWs, which have higher Seebeck coefficients than NWs in the other orientations (Fig. 2a). ii) More importantly, keeping $\sigma$ high, by utilizing light effective mass subbands, or using strain engineering to reduce the effective masses. In nanostructures band engineering is partially possible especially when utilizing devices in different orientations and, therefore, benefits can be achieved through proper optimization studies. iii) Possible ways to reach high carrier concentrations without direct doping, since this is such a strong performance degrading mechanism.

## V. Conclusions

The thermoelectric coefficients ($\sigma$, $S$, $\sigma S^2$, $k_e$, $ZT$) are calculated for n-type silicon NWs in different transport orientations for diameters $D$=3nm and $D$=12nm using the linearized Boltzmann approach. The $sp^3d^5s^*$-SO atomistic TB model was used for electronic structure calculation. Although, under ideal (ballistic) conditions, diameter scaling below 7nm can enhance the power factor and $ZT$ values of Si NWs by up to 2X [12], enhanced phonon scattering and especially SRS at those diameter scales weakens this possibility. Improved power factors in n-type Si NWs are, therefore, not obtained with diameter reduction, and depend on the changes and behavior of the conductivity rather than the Seebeck coefficient. Orientation can play a role in power factor optimization, with the [110] NWs having the best performance as the diameter scales to $D$=3nm because of their lighter mass subbands.

# Acknowledgements

This work was supported by the Austrian Climate and Energy Fund, contract No. 825467.

Figure 1:

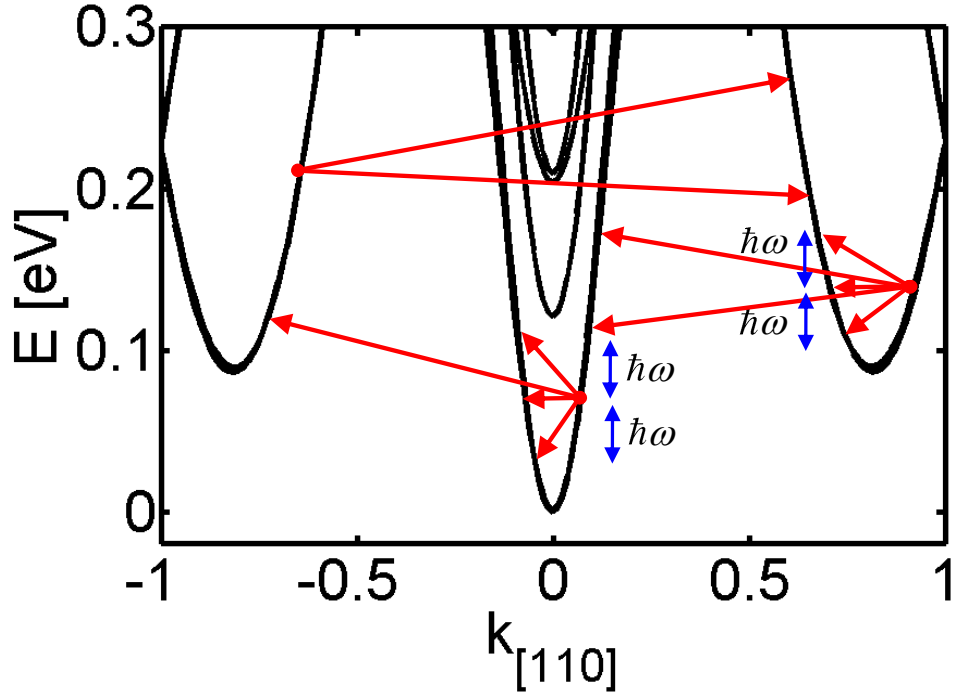

## Figure 1 caption:

Dispersion of the [110] NW with $D$=3nm with the scattering mechanisms indicated. Intravalley elastic and intervalley inelastic processes are considered (between the three valleys), following the bulk silicon scattering selection rules. Elastic processes: ADP, SRS and impurity scattering. Inelastic processes: Three for each *f*- and *g*- phonon scattering processes. Each valley in the dispersion is double-degenerate. Following the bulk scattering selection rules, however, each of the valleys is considered independently. $k_{[110]}$ is normalized to $2\pi/a_0'$, where $a_0' = 0.543/\sqrt{2}$ is the length of the unit cell in the [110] direction.

Figure 2:

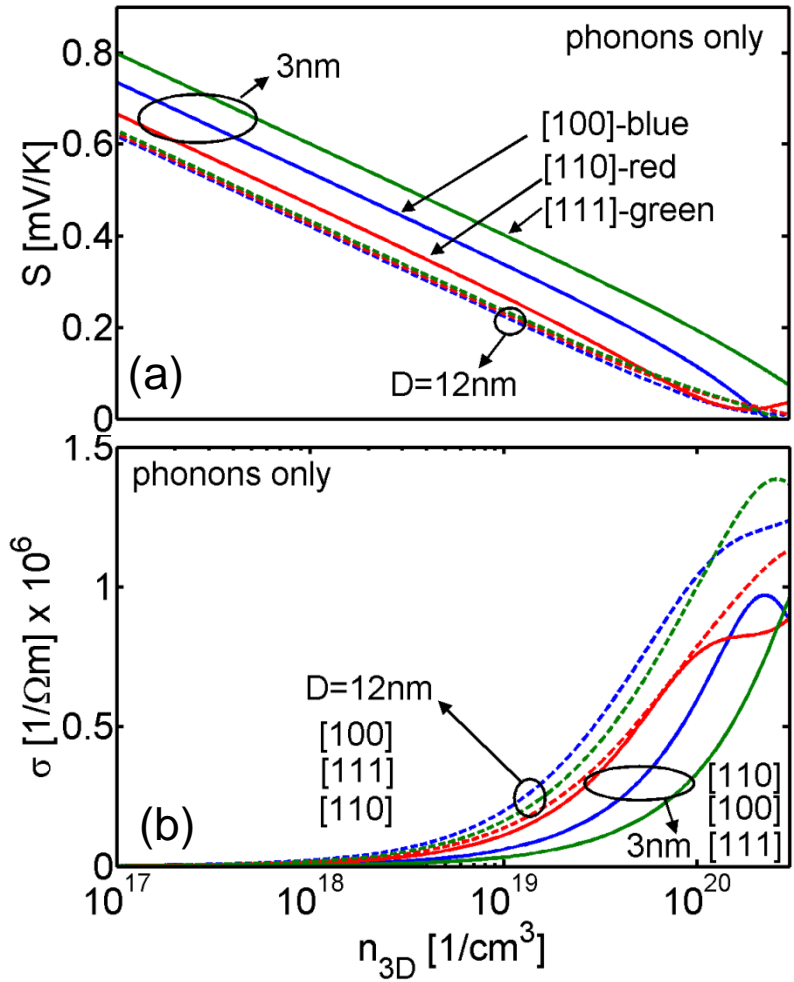

Figure 2 caption:

Thermoelectric coefficients for NWs in [100] (blue), [110] (red) and [111] (green) transport orientations and diameters $D$=3nm (solid) and $D$=12nm (dashed). (a) The Seebeck coefficients. (b) The electrical conductivity.

Figure 3:

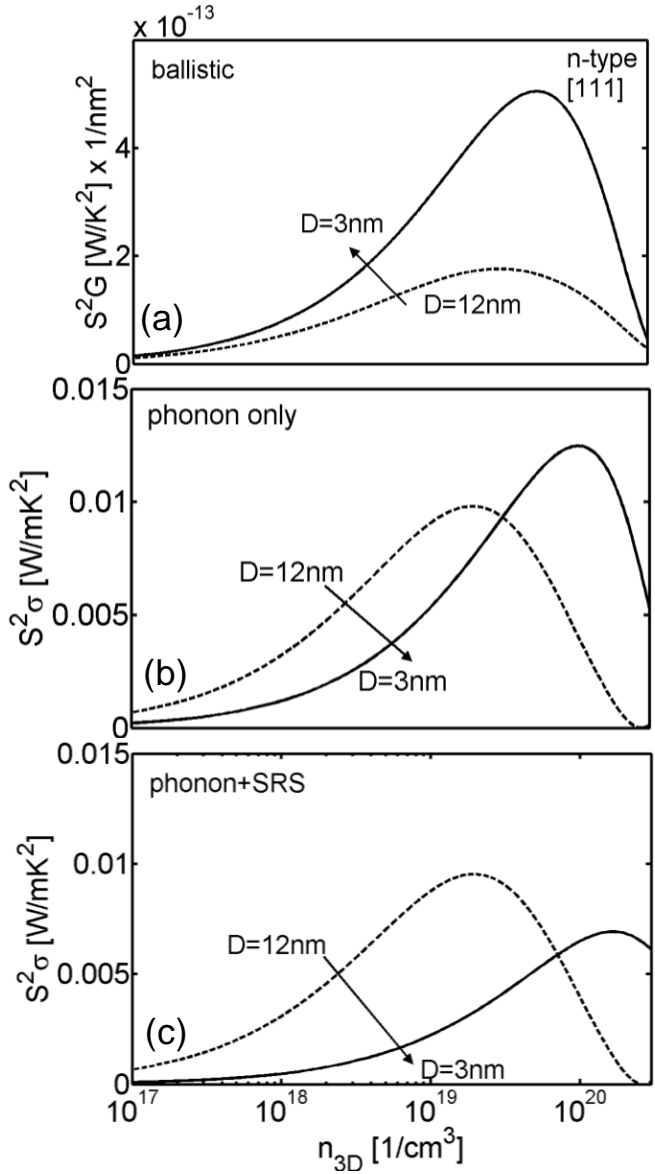

## Figure 3 caption:

The thermoelectric power factor for NWs in [111] transport orientation for diameters $D$=3nm (solid), $D$=12nm (dashed). (a) Ballistic transport conditions are considered. (b) Only phonon scattering is considered. (c) Phonon scattering and surface roughness scattering are considered.

Figure 4:

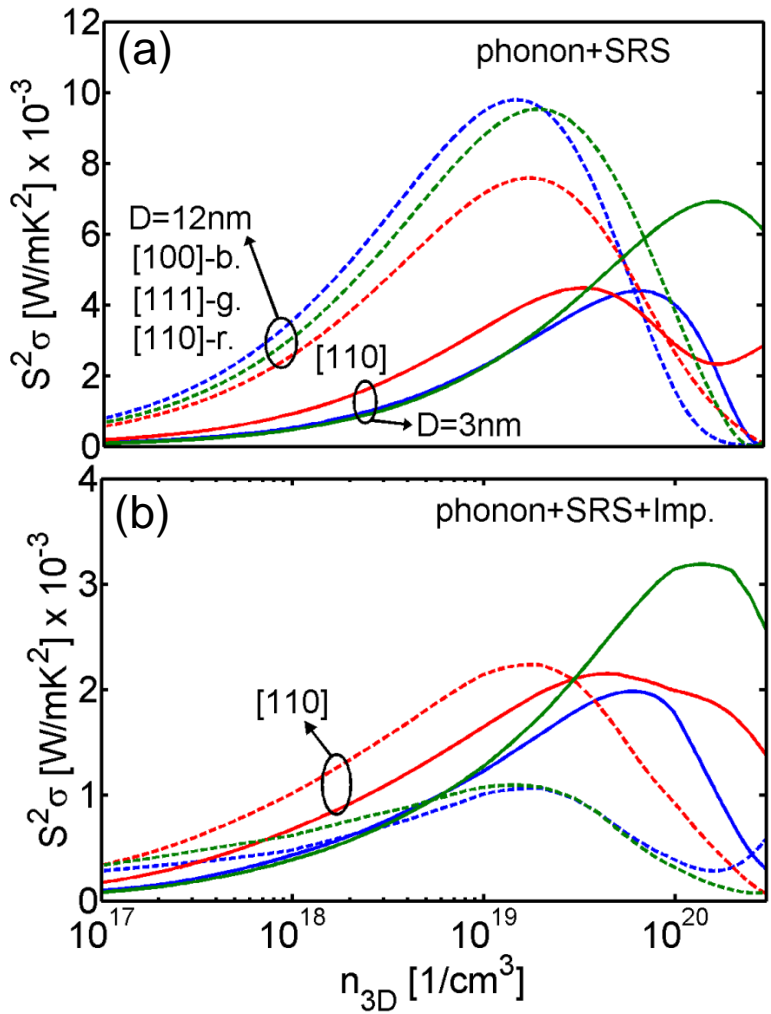

Figure 4 caption:

The thermoelectric power factor for NWs in [100] (blue), [110] (red) and [111] (green) transport orientations and diameters $D$=3nm (solid) and $D$=12nm (dashed). (a) Phonon and SRS are considered. (b) Phonon, SRS and impurity scattering are considered. The impurity concentration is equal to the carrier concentration.

Figure 5:

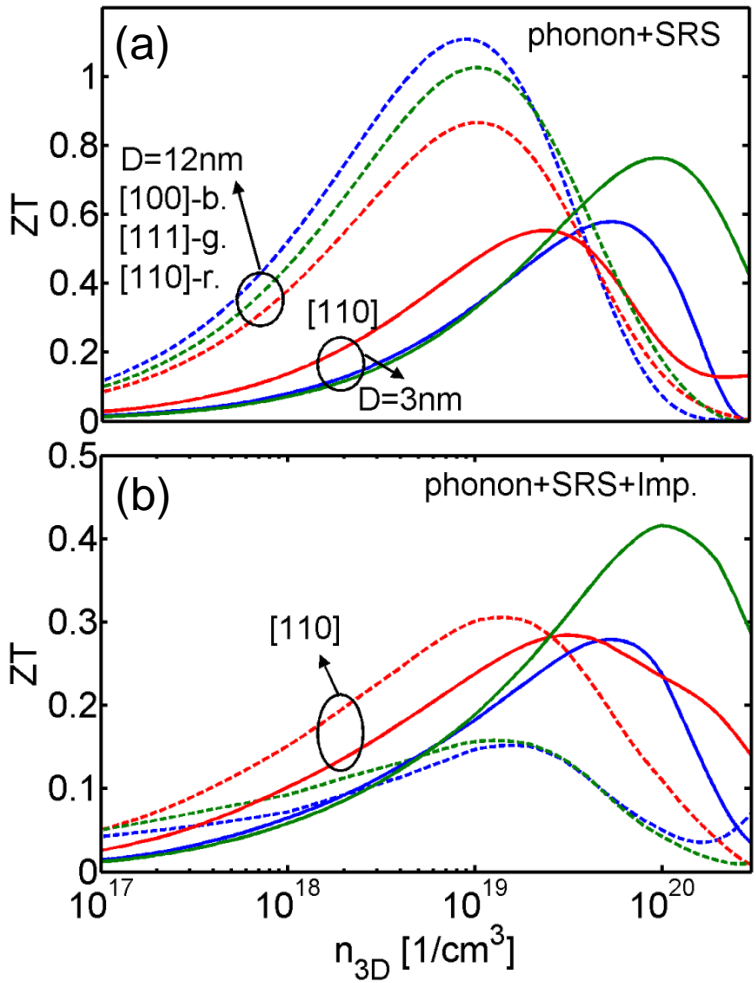

Figure 5 caption:

The thermoelectric figure of merit *ZT* for NWs in [100] (blue), [110] (red) and [111] (green) transport orientations and diameters *D*=3nm (solid) and *D*=12nm (dashed). The lattice part of the thermal conductivity is set to $k_l$=2W/mK in all cases. (a) Phonon and SRS are considered. (b) Phonon, SRS and impurity scattering are considered. The impurity concentration is equal to the carrier concentration.